\documentclass[aps, prx, twocolumn, showpacs, floatfix,10pt,superscriptaddress]{revtex4-2}
\usepackage{amsmath}
\usepackage{amsfonts}
\usepackage{amsbsy}
\usepackage{amssymb}
\usepackage{graphicx}
\usepackage{textcomp}
\usepackage[caption=false,singlelinecheck=false]{subfig}
\usepackage{xcolor}
\usepackage{mathrsfs}
\usepackage{mathtools}
\usepackage{bm}
\usepackage{gensymb}
\usepackage{braket,wasysym}
\usepackage[colorlinks,linkcolor=blue,citecolor=red,filecolor=magenta,urlcolor=red,breaklinks]{hyperref}
\usepackage[bottom]{footmisc}
\usepackage{verbatim}
\usepackage{bibunits}

\hypersetup{colorlinks=true, urlcolor=blue, citecolor=magenta, pdfborder={0 0 0}}
\usepackage{breakurl}
\usepackage{natbib}

\defaultbibliography{my1}

\allowdisplaybreaks

\normalfont
\def\be{\begin{equation}}
	\def\ee{\end{equation}}
\def\bea{\begin{eqnarray}}
	\def\eea{\end{eqnarray}}

\begin{document}
\title{Unveiling unique properties of icosahedral magnetic quasicrystals - Multipole physics and frustration}

\author{Junmo Jeon}
\email{junmo1996@kaist.ac.kr}
\affiliation{Korea Advanced Institute of Science and  Technology, Daejeon 34141, South Korea}
\author{SungBin Lee}
\email{sungbin@kaist.ac.kr}
\affiliation{Korea Advanced Institute of Science and  Technology, Daejeon 34141, South Korea}

\date{\today}
\begin{abstract}
Multipolar degrees of freedom and their hidden orders have been widely discussed in the context of heavy fermions, frustrated magnets and exotic Kondo effects. Although there has been extensive search for multipolar degrees of freedom in magnetic systems, there are few examples that allow pure multipolar degrees of freedom in the absence of magnetic dipoles. 
In this work, for the first time, we show that the magnetic behavior in an icosahedral quasicrystal is generally described by multipolar degrees of freedom, and in a specific case by the pure magnetic octupoles in the absence of dipoles, resulting from the interplay of spin orbit coupling and crystal field splitting. Importantly, we point out the non-crystallographic symmetries lead to multipolar degrees of freedom, only allowed in quasicrystals but forbidden in conventional crystals. For both Kramers and non-Kramers doublets, the characteristics of multipoles are classified and the effective spin Hamiltonian on symmetry grounds are derived. Based on the self-similar triangular structure of the icosahedron, we argue the long-range frustration in terms of the Ising model. We further classify the possible quantum phases including quantum fluctuations, in terms of the instantaneous entanglement generation of the ground state. 
Our study offers the magnetic icosahedral quasicrystal as a new platform to search for the novel multipolar degrees of freedom and their exotic phenomena.

\end{abstract}

\maketitle

\section{Introduction}\label{sec: intro}

In condensed matter systems, there are several examples that cannot be easily observed by conventional experimental techniques. Such {\it hidden} orders have been in debate for several decades and have been waiting to be discovered\cite{buyers1996low, shah2000hidden, chakravarty2001hidden,mydosh2011colloquium,mydosh2014hidden,doi:10.1021/jacs.1c09954}. 
In particular, unusual higher rank multipole moments, beyond the conventional electric and magnetic dipole moments, have been suggested as a key player to exhibit various hidden orders\cite{erkelens1987neutron, kitagawa1996possible, shiina1997magnetic, lingg1999ultrasound, yamauchi1999antiferroquadrupolar, tanaka1999evidence, kitagawa2000third,tayama2001antiferro, iwasa2002crystal, matsuoka2001evidence, buschow2003handbook,  hotta2005multipole,kiss2005group,  yanagisawa2005dilatometric, kiss2006scalar, santini2006hidden,tanida2006possible, matsuoka2008simultaneous, carretta2010quadrupolar, cricchio2009itinerant,  kusunose2011hidden,   ano2012quadrupole, onimaru2011antiferroquadrupolar, goho2015electronic, onimaru2016exotic, suzuki2018first}. 
Multipolar degrees of freedom are not only famous for their ability to give rise to hidden orders, but also for their role in driving a variety of other interesting and complex phenomena. 
For example, in heavy fermion materials, multipolar degrees of freedom can lead to the emergence of unconventional superconductivity and non-Fermi liquid behavior with exotic Kondo physics\cite{cox1987quadrupolar, tsvelik1993phenomenological,tou2005sb,walker1993nature, bauer2002superconductivity,kohgi2003evidence,  suzuki2005quadrupolar,kotegawa2008anomalous,  sakai2011kondo,  tsujimoto2014heavy,sumita2016superconductivity, onimaru2016quadrupole}. Moreover, beyond their hidden orders, magnetic frustration between   
 multipole moments can give rise to the emergence of exotic ground states, so called multipolar quantum spin liquids\cite{gao2019experimental,gaudet2019quantum,sibille2020quantum,PhysRevB.61.476}.
Hence, understanding the properties of multipolar physics have been the focus of intense research with broad implications and new insights with potential applications\cite{totsuka1995matrix, aoki2005novel, santini2009multipolar,tazai2019multipole, mydosh2020hidden,patri2020theory}. 
 However, there are limited examples of this kind and many of the magnetic systems contain not only multipoles but also magnetic dipoles at the same time. 
 
Finding multipolar degrees of freedom in the magnetic systems requires a delicate combination of spin-orbit couplings and crystalline electric field (CEF) splitting based on the point group symmetries\cite{cohen2016fundamentals,suzuki2018first}. In conventional crystals, the point group symmetry, which should be compatible with the translational symmetry, restricts searching pure multipolar degrees of freedom in the magnetic systems\cite{cohen2016fundamentals}. On the other hand, the quasicrystals, and their approximants could exhibit the point group symmetries beyond the space group such as pentagonal rotational symmetry because they are ordered without spatial periodicity\cite{kellendonk2015mathematics,suck2013quasicrystals,labib2020magnetic}. Thus, the quasicrystalline materials would be a good platform for finding multipolar degrees of freedom. Especially, several rare-earth magnetic quasicrystals are present having icosahedral symmetry but have never been explored in terms of their multipolar physics and related exotic phenomena\cite{deguchi2012quantum,suzuki2021magnetism,takeuchi2023high,de2007lattice,judd1957crystal,kashimoto1998magnetic}.

In this paper, we consider the noncrystallographic 5-fold rotational symmetry and icosahedral symmetry with $f$-orbital electrons of the rare earth atoms. 
First, we classify all possible multipole degrees of freedom present in rare-earth magnetic quasicrystals in the presence of noncrystallographic 5-fold  symmetry. We note that it generally leads to the higher order multipoles of the pseudospin $xy$ components, and magnetic dipoles of the  Ising moments.  More interestingly, if magnetic quasicrystals with Yb$^{3+}$ ions are in a perfect icosahedral crystal field symmetry, they host the Kramers doublet that carries pure magnetic octupole moments without magnetic dipole moments. On symmetry grounds, we introduce the generic spin Hamiltonian for both Kramers doublet and non-Kramers doublet. In the antiferromagnetic Ising limit, we first discuss the degenerate ground state born of the geometrically frustrated icosahedron structure. When the quantum fluctuation is introduced, we discuss the unique ground state is stabilized having  
non-zero entanglement. In this case, depending on (anti-) ferromagnetic $XY$ interaction, the specific linear combination of the degenerate ground states found in the Ising limit becomes a unique ground state. Our work provides a perspective for finding multipolar degrees of freedom and their magnetic frustration originated from noncrystallographic symmetries. Furthermore, it opens a new paradigm for enriching hidden orders, spin liquids and novel Kondo effects in quasicrystals.

\section{Results and discussion}
\subsection{Multipolar degrees of freedom in icosahedral magnetic quasicrystals}
\label{sec:model}
In this section, we first classify all possible multipolar degrees of freedom for  rare-earth magnetic systems in the presence of noncrystallographic 5-fold symmetry and icosahedral symmetry. Then, we focus on a special case where pure magnetic octupole moments exist in the absence of magnetic dipoles,  and discuss their characteristics. 

The most general CEF Hamiltonian in a 5-fold rotational symmetry is given as follows, using the Stevens operators,
\begin{align}
\label{CEFH0}
&H_{\mbox{CEF}}=B_{60}O_{6}^{0}+B_{65}O_{6}^{5}+B_{20}O_{2}^{0}+B_{40}O_{4}^{0}
\end{align}
Here, $B_{nm}=-\gamma_{nm}qC_{nm}\braket{r^n}\theta_n$ are the Stevens coefficients obtained by the radial integrals, where $\gamma_{nm}$ is a term calculated from the ligand environment expressed in terms of tesseral harmonics. $q$ is the charge of the central atom, $C_{nm}$ are the normalization factors of the spherical harmonics, $r$ is the radial position, $\theta_n$ are constants associated with electron orbitals of the magnetic ion\cite{Scheie:in5044}. Here, we assume the point charge model. Note that the $z$-axis is a 5-fold rotational symmetry axis. The $O_n^m$ are Stevens operators with respect to the total angular momentum operators (See Supplementary Materials for the detailed forms of $O_n^m$.). When we consider the full icosahedral symmetry, which has 5-fold rotational symmetry plus mirror reflection symmetry, it leads to $B_{20}=B_{40}=0$, and $B_{65}=-42B_{60}$. 
We consider both cases where a perfect icosahedral symmetry is present or only 5-fold symmetry is present, applicable to different situations respectively. The latter case can be taken into account using the charge defects which breaks the icosahedral symmetry down to $C_{5v}$. 
In this case, the charge defect on a ligand is placed on a $z$-axis as $q\to q(1+\alpha)$. 
 As a result, the Stevens coefficients, $B_{65},B_{20},B_{40}$ are changing as a function of $\alpha$. Note that $B_{nm}=-\alpha\gamma_{nm}'qC_{nm}\braket{r^n}\theta_n$ for a single point charge defect,  $B_{60}$ is unchanged if the charge defect is placed on the $z$-axis.

\begin{figure*}
\centering
  \includegraphics[width=0.95\textwidth]{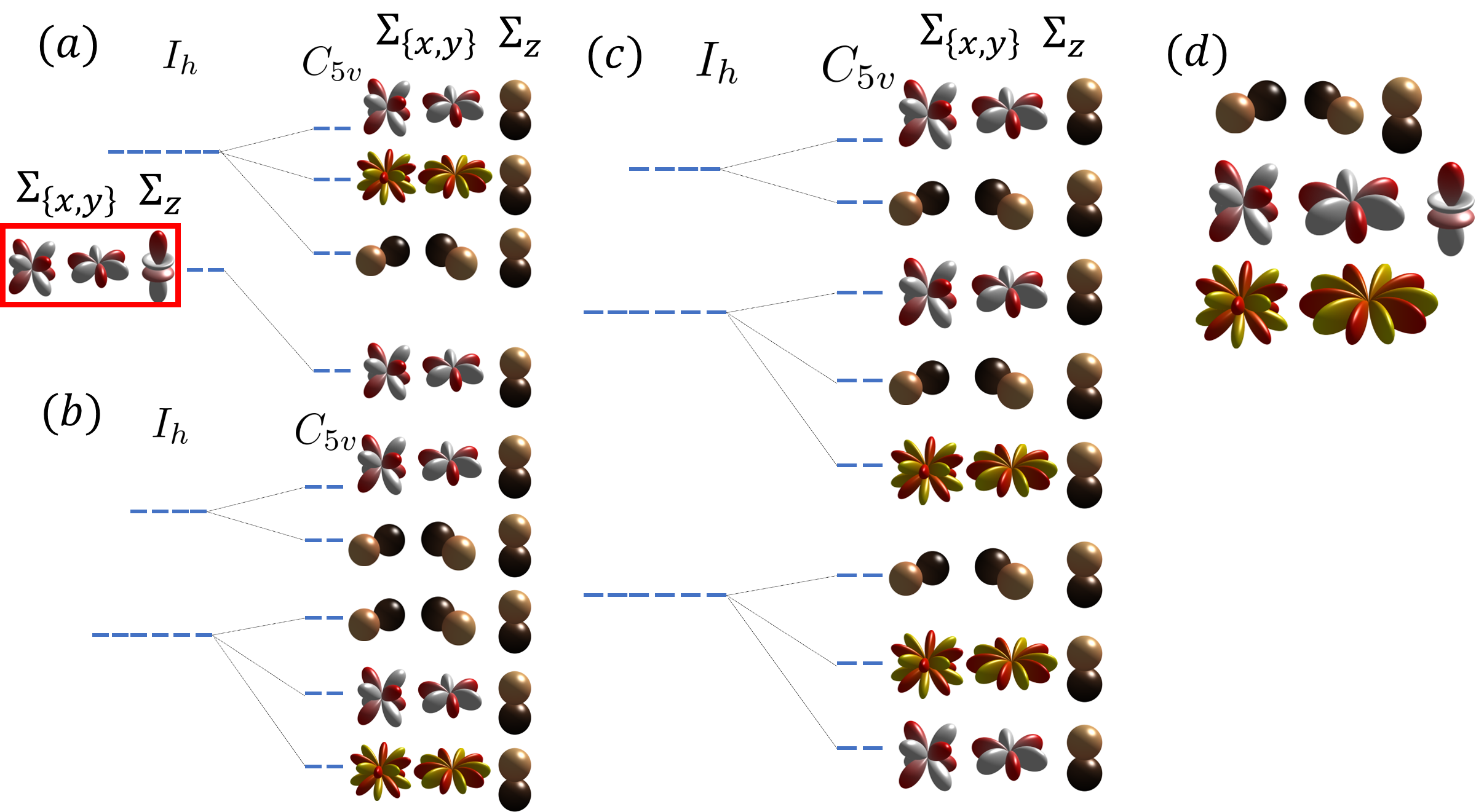}
  \caption{Summary of the CEF states and multipoles under noncrystallographic point group symmetries for given half-integer values of $J$, (a) $J=7/2$ (Yb$^{3+}$) (b) $J=9/2$ (Nd$^{3+}$) and (c) $J=15/2$ (Dy$^{3+}$) respectively. Under the icosahedral symmetry $I_h$, the CEF states are split into several multiplets. 
Particularly, note that $J=7/2$ which is the case for Yb$^{3+}$ has a unique Kramers doublet and it hosts pure magnetic octupoles without magnetic dipoles or quadrupoles, as shown in the red box. Whereas, for $C_{5v}$ where the mirror reflection is absent from $I_h$, all the multiplets are split into either Kramers doublets. In this case, the $z$-components of the pseudospin, $\Sigma_z$, are dipoles for every doublet state. But, the $xy$-components, $\Sigma_{x,y}$, are not only dipoles but also octupoles and dotriacontapoles. Depending on the energy scale of the breaking of mirror symmetry, the order of energy could change. (d) Dipoles (black-copper), octupoles (red-white) and dotriacontapoles (red-yellow). See the main text for more details.}
  \label{fig: halfint}
\end{figure*}

\begin{figure*}
\centering
  \includegraphics[width=0.95\textwidth]{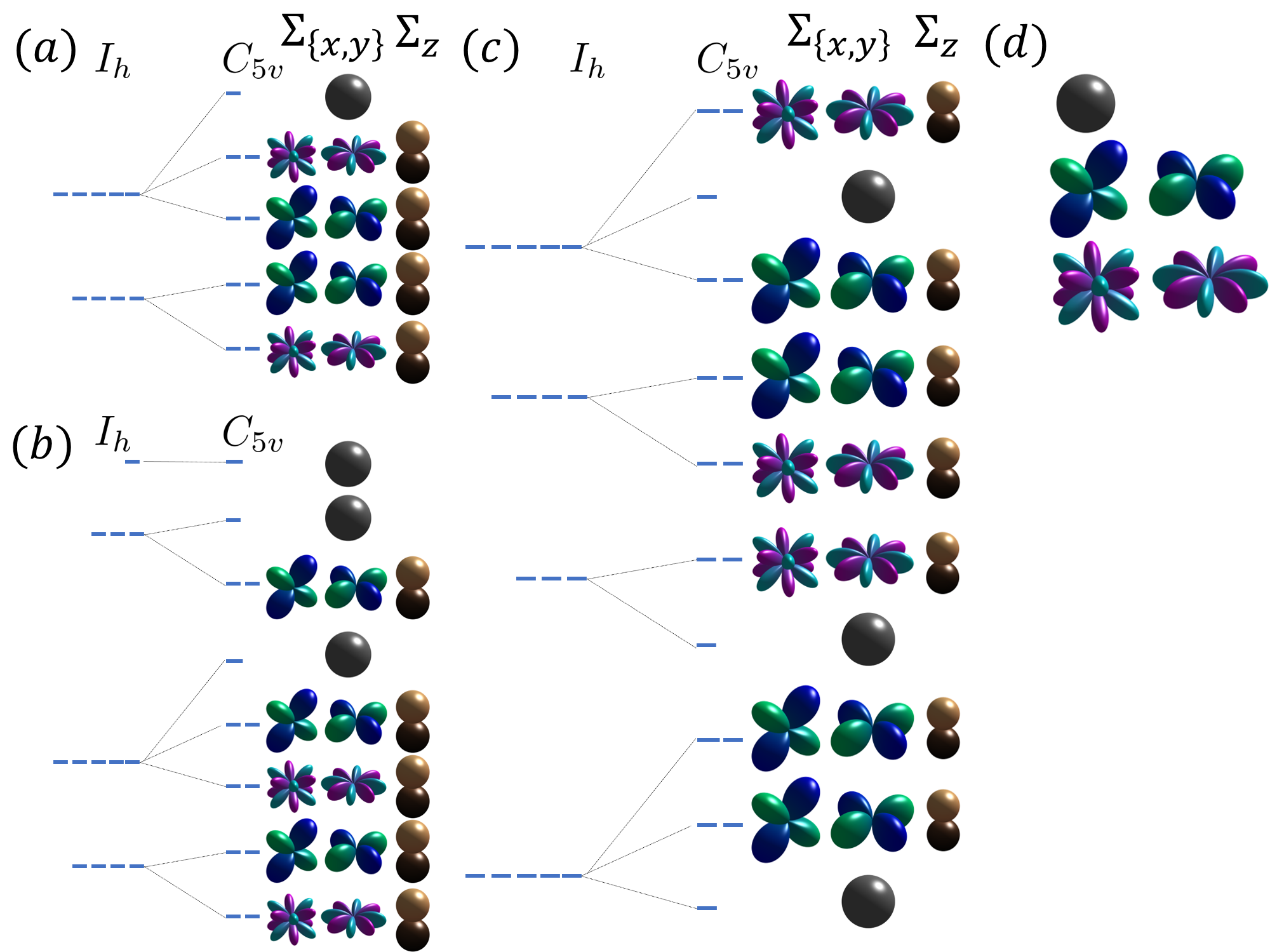}
  \caption{ Summary of the CEF states and multipoles under noncrystallographic point group symmetries for given integer values of $J$, (a) $J=4$ (Pr$^{3+}$) (b) $J=6$ (Tb$^{3+}$) and (c) $J=8$ (Ho$^{3+}$). 
 Unlike the icosahedral symmetry $I_h$ where the CEF states are split into several multiplets. the five-fold symmetry $C_{5v}$ symmetry splits all the multiplets into either non-Kramers doublets or singlets.  
     Under $C_{5v}$, the $z$-components of the pseudospin, $\Sigma_z$ are magnetic dipoles for every doublets. But, the $xy$-components, $\Sigma_{x,y}$ represent either quadrupoles or hexadecapoles. Depending on the energy scale of the breaking of mirror symmetry, the order of energy could change.  (d) Singlet (black), quadrupoles (blue-skyblue) and hexadecapoles (violet-skyblue). See the main text for more details.}
  \label{fig: int}
\end{figure*}

For both full icosahedral symmetry, $I_h$ and 5-fold rotational symmetry, $C_{5v}$ with nonzero $\alpha$, we summarize the CEF states in Figs.\ref{fig: halfint} and \ref{fig: int} for half-integer $J$ and integer $J$ values, respectively. 
For $\alpha=0$, the point group symmetry restores the full icosahedral symmetry and multiple degenerate levels appear. It is noteworthy that for $J=7/2$ which is the case for Yb$^{3+}$, the unique (Kramers) doublet exists under the perfect icosahedral symmetry group as shown in red box of Fig.\ref{fig: halfint} (a). Specifically, there are two eigenspaces of $H_{\mbox{CEF}}$, the Kramers doublet and the sextet\cite{walter1987crystal}.
On the other hand, for any given $\alpha\neq 0$, the CEF states under 5-fold symmetry are split into doublets or singlets for all given values of the total angular momentum $J$ of the rare earth atom. In both Fig.\ref{fig: halfint} and Fig.\ref{fig: int}, the tentative orders of the energy levels are given for $\alpha=0.5$. Given $\alpha\neq 0$, every energy levels for half-integer $J$ are Kramers doublet due to the time-reversal symmetry. However, for integer $J$, some singlets are allowed. Notably, for every case, the $z$-component of the pseudospins of the doublets are magnetic dipoles, while the $x$ and $y$-components of the pseudospins of the doublets are generally multipoles such as quadrupoles, octupoles, and even higher ones. For half-integer values of $J$, $\Sigma_{x(y)}$ could be dipole, octupole and dotriacontapole (see Fig.\ref{fig: halfint}), whereas for integer values of $J$, the doublets carry either quadrupole or hexadecapole $\Sigma_{x(y)}$ (see Fig.\ref{fig: int}). Again, it is because of the time-reversal symmetry. It is interesting to note that for the low lying Kramers doublet, $\Sigma_{x(y)}$ for $J=7/2$ always represents magnetic octupoles originated from the unique Kramers doublet, whereas, $\Sigma_{x(y)}$ for $J=9/2$ and $J=15/2$ could take magnetic octupoles or dotriacontapoles. We also note that for the low lying non-Kramers doublets, $\Sigma_{x(y)}$ for $J=4$ and $J=6$ take either quadrupole or hexadecapole, while for $J=8$, it could take only quadrupole (see Fig.\ref{fig: int} (c)).


From now on, we focus on the special case where the lowest Kramers doublet is represented by pure octupoles, {i.e.}  $J=7/2$  in an icosahedral crystal field symmetry and discuss their characteristics.
Then, in the following subsections, we discuss the generic spin Hamiltonian, geometrical frustration and quantum fluctuation effects. It is important to note that all these arguments are generally applicable to any cases of multipolar physics with $J$ listed in Fig.\ref{fig: halfint}  and Fig.\ref{fig: int}. 

Let us define the Kramers doublet as $\left\vert {\pm}\right\rangle$, which are written in terms of the eigenstates of the $J_z$ operator,
\begin{align}
\label{Kramers}
&\left\vert{+}\right\rangle=-\sqrt{\frac{7}{10}}\left\vert{J_z\!\!=\!\!-\frac{3}{2}}\right\rangle+\sqrt{\frac{3}{10}}\left\vert{+\frac{7}{2}}\right\rangle \\
&\left\vert{-}\right\rangle=\sqrt{\frac{3}{10}}\left\vert{-\frac{7}{2}}\right\rangle+\sqrt{\frac{7}{10}}\left\vert{+\frac{3}{2}}\right\rangle. \nonumber
\end{align}
In case of Yb$^{3+}$ ion, it is known that $B_6<0$, so the Kramers doublet is the ground eigenspace of the CEF Hamiltonian, well separated from the sextet. \cite{lindgard1975tables,sandberg2021emergent}. 
The CEF energy gap is given by $25200 \vert B_{60}\vert\sim$ $\mathcal{O}$ (10meV), where detailed number depends on icosahedral magnetic materials with Yb. 
Since the energy scale of spin exchanges between rare-earth ions are generally much smaller than the crystal field splittings, one can expect the magnetic properties at low temperature are explained within this Kramers doublet. 
 From Eq.\eqref{Kramers}, one can easily find that $\braket{\pm\vert J_i\vert\mp}$ and $\braket{\pm\vert J_i\vert\pm}$ vanish where $i=x,y,z$. Importantly, $J_z$ also vanishes due to the symmetric coefficients of the states. This confirms that there is no magnetic dipole moment. Thus, one should consider the multipolar degrees of freedom given by the irreducible tensor operators. However, since $\left\vert{\pm}\right\rangle$ are the Kramers doublet, the time-reversal even operators such as quadrupole moments vanish. Hence, we can expect the higher order degrees of freedom such as octupoles, in the absence of any dipolar or quadrupolar degrees of freedom.

To show that the Kramers doublet, $\left\vert{\pm}\right\rangle$ in Eq.\eqref{Kramers}, describes the octupolar degrees of freedom, let us define the pseudospin ladder operators, $\Sigma^\pm$, as follows.
\begin{align}
\label{effectiveladder}
&\Sigma^+=\left\vert{+}\right\rangle \left\langle{-}\right\vert \ \ \ \ \ \Sigma^-=(\Sigma^+)^{\dagger},
\end{align}
and $\Sigma^{z}=[\Sigma^+,\Sigma^-]/2$. Now define the octupolar operators as the rank 3 spherical tensor operators, $T_m^{(3)}$ in terms of $J_+,J_-$ and $J_z$. 
Note that octupolar operators are time-reversal odd, under the time-reversal transformation, $\mathcal{T}$, it satisfies
$\mathcal{T}\Sigma^\pm\mathcal{T}^{-1} \!\!=\!\! -\Sigma_\mp$, $\mathcal{T}\Sigma^{z}\mathcal{T}^{-1} \!\!=\!\!-\Sigma^{z}$.
As a result, $\Sigma^{z}\sim T_0^{(3)}$ and $\Sigma^\pm\sim T_m^{(3)}$ for non-zero $m$. However, $T_1^{(3)}$ and $T_{-1}^{(3)}$ vanish because $T_{\pm 1}^{(3)}\left\vert{\pm}\right\rangle$ is not in the doublet eigenspace. Note that $T_{\pm 1}^{(3)}$ changes the eigenvalue of the $J_z$ operator by $\pm 1$. Similarly, since $J_\pm^{2}\left\vert{\pm}\right\rangle$ and $J_\pm^3\left\vert{\mp}\right\rangle$ are not in the doublet, the only non-trivial matrix elements are $\braket{\pm\vert T_{\pm 2}^{(3)}\vert\mp}$ and $\braket{\mp\vert T_{\pm 3}^{(3)}\vert\pm}$. This leads to $T_{\pm 2,\mp 3}^{(3)}\sim \Sigma^\pm$. In detail, one can represent the octupole pseudospin operators in the doublet as,
\begin{align}
\label{octupoles}
\Sigma^{z}\equiv T_0^{(3)},  \ \ \ &\Sigma^\pm\equiv \frac{1}{2}\sqrt{\frac{2}{15}}T_{\pm2}^{(3)}\mp\frac{1}{2}\sqrt{\frac{1}{5}}T_{\mp 3}^{(3)} .
\end{align}
Specifically, $T_{\pm2}^{(3)}=\frac{1}{4}\sqrt{\frac{105}{\pi}}\overline{J_\pm^2J_z}$, $T_{\pm3}^{(3)}=\mp\frac{1}{8}\sqrt{\frac{35}{\pi}}J_\pm^3$, and $T_0^{(3)}=\frac{1}{4}\sqrt{\frac{7}{\pi}}(5J_z^3-3J_zJ^2)$, where $\overline{\mathcal{O}}$ is the symmetrization of the operator $\mathcal{O}$\cite{sakurai1995modern}. Each pseudospin operator behaves like the rank 3 tensors. From Eq.\eqref{octupoles}, one can write,
\begin{align}
\label{xyspinop}
&\Sigma_{x(y)}\!=\! \frac{1}{4}\left[ \sqrt{\frac{2}{15}}(T_2^{(3)}\pm T_{-2}^{(3)})\pm \sqrt{\frac{1}{5}}(T_3^{(3)}-T_{-3}^{(3)})\right],
\end{align}
where $x(y)$ takes $+(-)$ sign in Eq.\eqref{xyspinop}, respectively.

\subsection{Generic spin Hamiltonian on symmetry grounds}
\label{subsec: spin-Hamiltonian}
By applying the symmetry transformation of the icosahedron group ($I_h$) and the time reversal symmetry transformation, one can find the generic spin Hamiltonian of the nearest neighbor interactions. Let us define the local $z$-axis pointing to the center of the icosahedron shell. Then, under the 5-fold rotational symmetries of $I_h$, we have $\Sigma^\pm_{i}\to e^{\mp 4i\pi/5}\Sigma^\pm_{j}$, where $i$ and $j$ are the nearest neighbor sites. This leads to the bond dependent phase factors in the Hamiltonian, such as $\Sigma^+_{i}\Sigma^+_{j}$ or $\Sigma^+_{i}\Sigma^{z}_{j}$. As a result, the generic symmetry allowed Hamiltonian under the icosahedral point group symmetry contains the four independent parameters, $J_{\pm\pm},J_{\pm z},J_{\pm}$ and $J_{zz}$, and is given as, 
\begin{align}
\label{H}
H&=\sum_{\braket{i,j}} 
\Bigl[  J_{zz}\Sigma_{i}^z\Sigma_{j}^z 
+  J_{\pm}  \big(\Sigma_{i}^+\Sigma_{j}^- +\Sigma_{i}^- \Sigma_{j}^+ \big) \\
&+ J_{\pm\pm} \big( \alpha_{ij} \Sigma_{i}^+\Sigma_{j}^+ + \alpha_{ij}^* \Sigma_{i}^-\Sigma_{j}^- \big)  \nonumber \\
&+ J_{\pm z}  \Big( \Sigma_{i}^z \big(\beta_{ij} \Sigma_j^+ +\beta_{ij}^* \Sigma_j^- \big) + i \leftrightarrow j  \Big) \nonumber
\Bigr].  
\end{align}
Here, $\alpha_{ij}$ takes the values $1,e^{\pm i 2\pi /5}$ and $e^{\pm i 4\pi /5}$ depending on the bond orientation due to the 5-fold rotational symmetry, and $\beta_{ij}=(\alpha_{ij}^*)^2$. The Hamiltonian in Eq.\eqref{H} is written in terms of local coordinate axes, where the local $z$-axis for each sites as pointing in the icosahedron (See Supplementary Materials for detailed derivation of the effective pseudospin Hamiltonian and local axes.). The magnitudes of these spin exchange parameters would vary from case to case. Here, for the simplest case, we first study the Ising limit with a finite $J_{zz}$ and then consider the quantum fluctuations in the presence of  $J_{\pm}$.

Remarkably, we should emphasize that the spin Hamiltonian, $H$ in Eq.\eqref{H}, could be used to argue general characteristics of the multipolar pseudospin models even for the $C_{5v}$ symmetry rather than $I_h$. The Hamiltonian in Eq.\eqref{H} is applicable for any Kramers or non-Kramers doublets with half-integer or integer values of $J$, in  the case of $C_{5v}$ under small charge defect on the \textit{ligand} which has been discussed in terms of $\alpha$. The only difference between Kramers and non-Kramers doublets is the presence and the absence of the parameter $J_{\pm z}$. This is originated from the fact that for non-Kramers doublet, $z$ components represent magnetic dipoles, whereas $x,y$ components represent quadrupoles or hexadecapoles, thus the coupling between $\Sigma^z$ and $\Sigma^\pm$ is forbidden under the time reversal symmetry.  

\begin{figure}
\centering
  \includegraphics[width=0.45\textwidth]{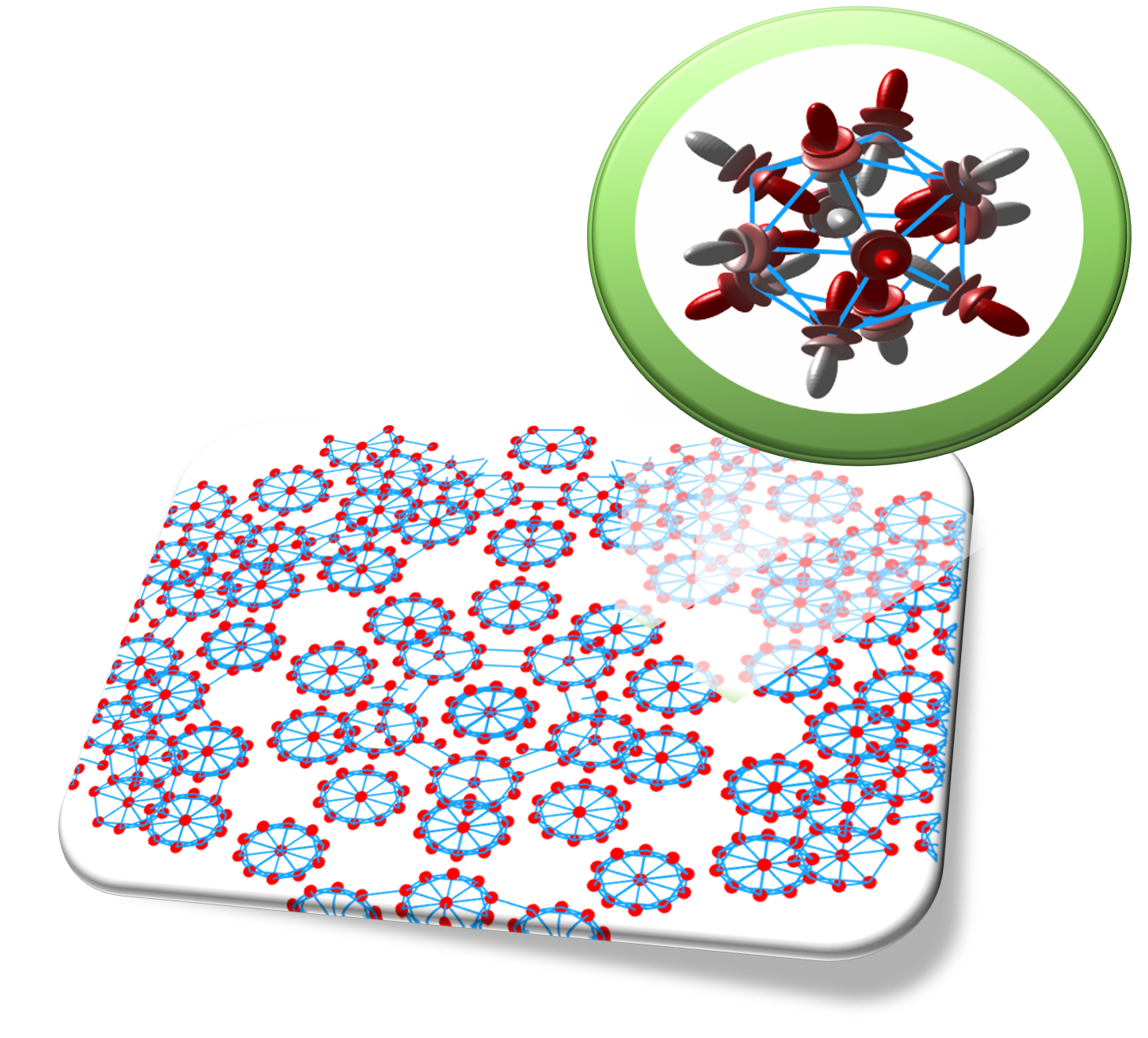}
  \caption{Icosahedral quasicrystal generated by the cut-and-project scheme from 6-dimensional hyperspace. We draw the sites forming the icosahedron shells only for visibility. The distances between neighboring icosahedron shells vary. For the antiferromagnetic, $J_{zz}$, there is geometrical frustration on each icosahedron.}
  \label{fig: main}
\end{figure}

\subsection{Geometrical frustration}
\label{sec:delocalization}
Let us first consider the Ising model, where only $J_{zz}$ is non-zero in Eq.\eqref{H}. Considering the icosahedral quasicrystal descended from 6-dimensional hyperspace\cite{kellendonk2015mathematics}, Fig.\ref{fig: main} represents the structure of icosahedral quasicrystal. As shown in Fig.\ref{fig: main}, the distances between inter-shell of icosahedron vary  (See Supplementary Materials for detailed cut-and-project scheme for the icosahedral quasicrystal.). In real materials, depending on the structures of quasicrystals and approximants, the distances between the shells could differ\cite{suzuki2021magnetism,de2007lattice,takeuchi2023high,steinhardt1987physics, deguchi2012quantum}.
Furthermore, it is known that the inter-shell distance can be also controlled in terms of the external pressure\cite{berns2013problem,deguchi2012quantum,watanuki2006pressure}.
 Thus, for general argument, we mainly focus on the nearest neighboring sites in a single icosahedron and discuss the magnetic states. Depending on the inter-shell distances, one may consider the perturbative approach (See Supplementary Materials for perturbative approach.). For ferromagnetic $J_{zz}$, it is obvious there are only two degenerate ground states exist, where every octupole points in local $+z$ and $-z$ directions, respectively.

For antiferromagnetic Ising, $J_{zz}>0$, the triangular faces of the icosahedron cause geometric frustration. In this case, 72 degenerate states exist and they are classified into two groups on symmetry grounds: (i) 60 degenerate states without 5-fold rotational symmetry, (ii) 12 degenerate states with 5-fold rotational symmetry.
\begin{figure}
\centering
  \includegraphics[width=0.45\textwidth]{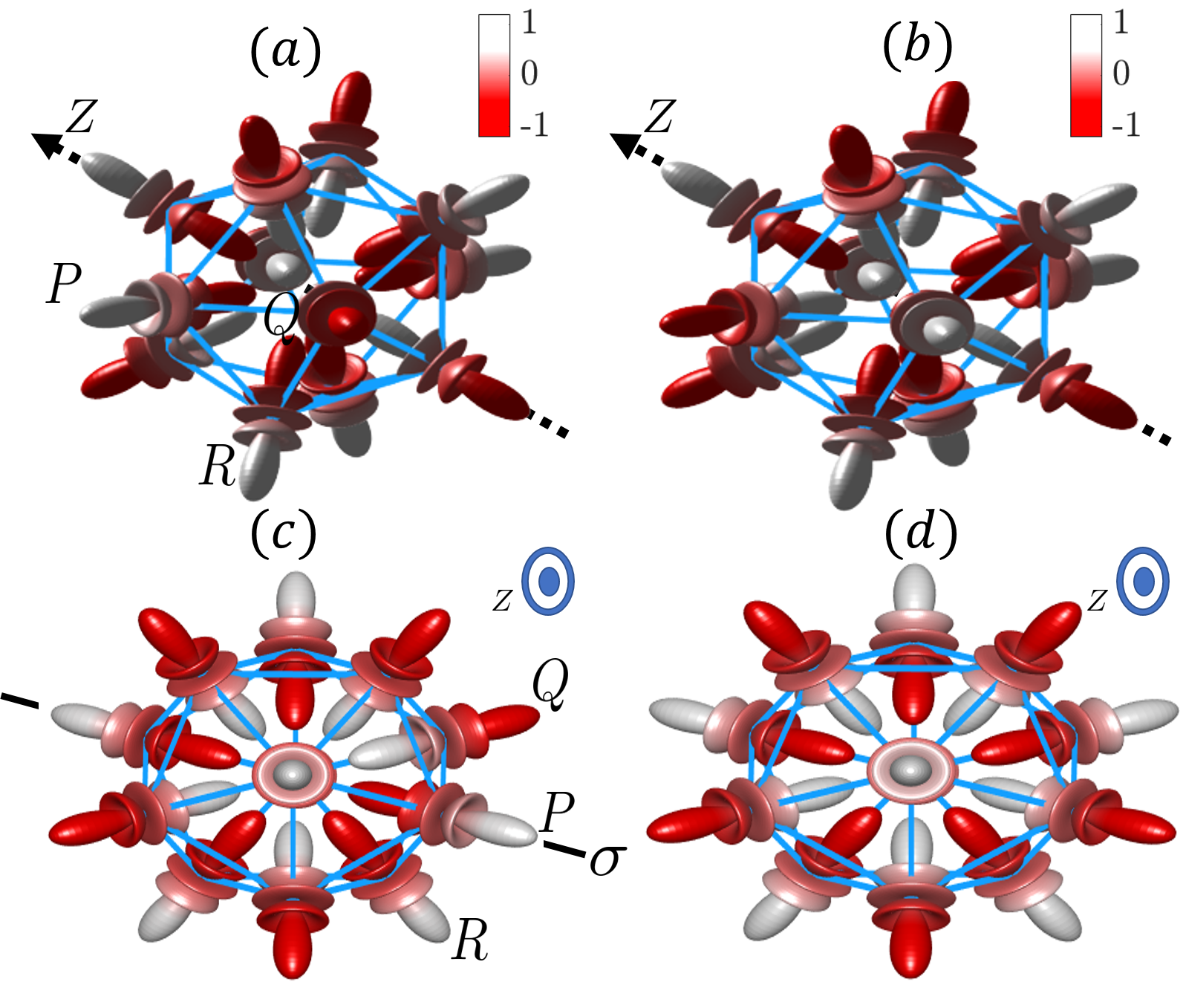}
  \caption{Two distinct octupolar ground states on an icosahedron for antiferromagnetic $J_{zz}$ based on the symmetry ground. (a) 60 degenerate states with no 5-fold rotational symmetry. (b) 12 degenerate states with 5-fold symmetry. $Z$ axis is the 5-fold rotational symmetry axis in (b). (c,d) Top view along $Z$ axis in (a,b), respectively. (c) Mirror reflection with respect to $\sigma$ plane containing $Z$ axis swaps the octupoles on the sites Q and R that leads to another degenerate state without 5-fold symmetry. (d) The state depicted in (b) possesses 5-fold rotational symmetry along $Z$ axis.}
  \label{fig: 72}
\end{figure}
Fig.\ref{fig: 72} (a) show an example of the first group of ground states. Note the octupolar moments arranged on the icosahedron in Fig.\ref{fig: 72} (a) do not have a 5-fold rotational symmetry axis. Since there are 6 independent choices of the $Z$ axis, by applying 5-fold rotational transformation around each $Z$ axis, we have 30 degenerate states. In addition, for each 30 degenerate states, the energy is invariant under the swap of two octupoles on the sites Q and R in Figs.\ref{fig: 72} (a) and (c). Hence $M_\sigma$, the spatial mirror reflection with respect to the $\sigma$-plane depicted in Fig.\ref{fig: 72} (c) doubles the number of degenerate states with no 5-fold rotational symmetry, resulting in total 60 degenerate states. (See Supplementary Materials for detailed discussion of the symmetry argument.). 
Figs.\ref{fig: 72} (b) and (d) illustrate 5-fold rotational symmetric ground state. There are 12 independent choice of the rotational symmetry axis, $Z$ axis in Fig.\ref{fig: 72} (b) of the second group.

\subsection{Quantum fluctuation}
\label{sec:Fibonacci}
Now let us consider non-zero but small $J_\pm$ and study the effect of quantum fluctuations. 
To study the fluctuation effects, we introduce three subsets of the 72 degenerate states, $A,B$ and $C$ in terms of the \textit{orientation preserving} icosahedral rotation group, $I\trianglelefteq I_h$. Specifically, the subsets $A$ and $C$ are generated by applying the spatial rotations in $I$ to the states in Figs.\ref{fig: 72} (a) and (b), respectively. While, the subset $B$ is generated by applying the coset, $IM_\sigma=\{gM_\sigma \vert g\in I\}$ to the state in Fig.\ref{fig: 72} (a). Thus, for $H_\pm=J_\pm\sum_{\braket{i,j}}(\Sigma_i^+\Sigma_j^-+\Sigma_i^-\Sigma_j^+)$, two states in the same subset admit zero matrix element of $H_\pm$.   
One can let $\left\vert{\psi_{A_n}}\right\rangle$, $\left\vert{\psi_{B_l}}\right\rangle$ and $\left\vert{\psi_{C_r}}\right\rangle$, where $1\le n,l\le 30$ and $1\le r\le 12$ be the states in $A,B$ and $C$.
 Hence, in the sub-Hilbert space of the 72 Ising ground states, $H_\pm$ has the matrix representation, $[H_\pm]_{A,B,C}$, given by
\begin{align}
\label{Hpm}
&[H_\pm]_{A,B,C}=\begin{pmatrix} 0 & T_{AB} & T_{AC} \\ T_{BA} & 0 & T_{BC} \\ T_{CA} & T_{CB} & 0 \end{pmatrix}
\end{align}
where $T_{BA}=T_{AB}^{\dagger}$ is a $30\times30$ matrix, while $T_{AC}=T_{CA}^{\dagger}$ and $T_{BC}=T_{CB}^{\dagger}$ are $30\times12$ matrices. Here, each non-zero matrix element is $J_\pm$. 
On symmetry grounds, we can write the general form of the ground state, $\left\vert{GS}\right\rangle$, as,
\begin{align}
\label{GS}
&\left\vert{GS}\right\rangle=a\sum_{n=1}^{30}\left\vert{\psi_{A_n}}\right\rangle+b\sum_{l=1}^{30}\left\vert{\psi_{B_l}}\right\rangle+c\sum_{r=1}^{12}\left\vert{\psi_{C_r}}\right\rangle,
\end{align}
where we have only three real coefficients, $a,b$ and $c$ for $\left\vert{\psi_{A_n}}\right\rangle$, $\left\vert{\psi_{B_l}}\right\rangle$ and $\left\vert{\psi_{C_r}}\right\rangle$, respectively
(See Supplementary Materials for detailed discussion for the perturbative method.). 
 The energy correction is $E(a,b,c)=\braket{GS\vert H_\pm\vert GS}$.

First, considering $J_\pm<0$, the Lagrange multiplier method leads to $a=b=(1+\sqrt{6})c/5$ for the ground state. Next, if $J_\pm>0$, $E(a,b,c)$ is minimized when $a=-b$ and $c=0$. Remarkably, we have no degeneracy in either cases. Thus, any small quantum fluctuation given by $H_\pm$ leads to a unique ground state with particular superpositions of degenerate states (See Supplementary Materials for detailed derivation).

\begin{figure}
\centering
  \includegraphics[width=0.45\textwidth]{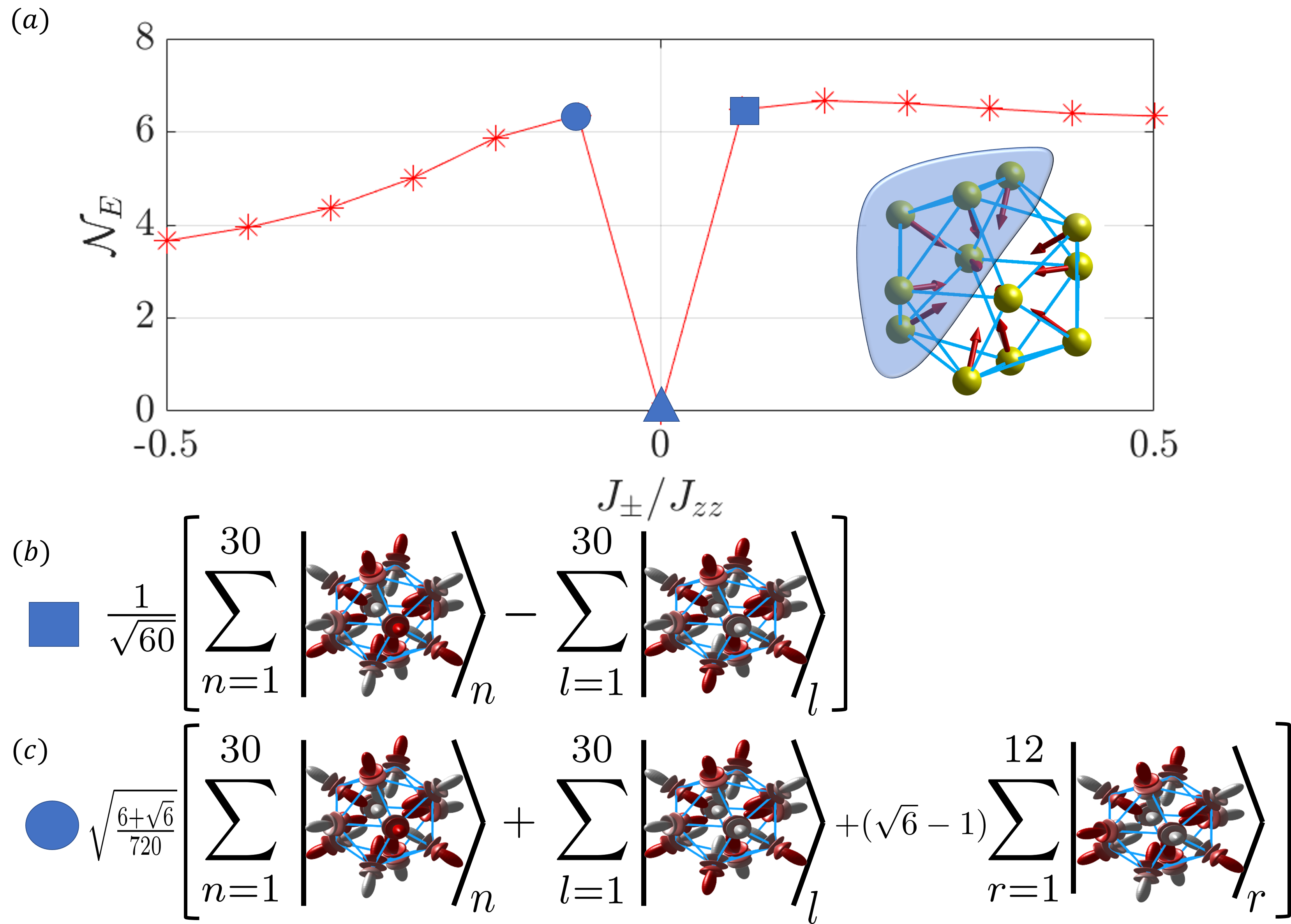}
  \caption{
Quantum fluctuation effect of nonzero $J_\pm$. (a) For nonzero $J_\pm$, the entanglement negativity of the ground state, $\mathcal{N}_E$, becomes finite. The sub-Hilbert space is defined by the blue shaded region of the icosahedron as shown in the inset. Red arrows are drawn to emphasize the local $z$ axes. At $J_\pm=0$ (marked by the triangle), the state is separable. For $J_\pm>0$ (square, (b)), the unique ground state is comprised of the states in the $A$ and $B$ subsets without 5-fold symmetry, with equal probability but opposite sign. For $J_\pm<0$ (circle, (c)), all 72 degenerate states are superposed in the unique ground state as shown in (c).
}
  \label{fig: entanglement}
\end{figure}

To capture the entanglement, we compute the entanglement negativity of the state defined by $\mathcal{N}_E=\sum_i(\left\vert\lambda_i\right\vert-\lambda_i)/2$, where $\lambda_i$ are all eigenvalues of $\rho^{T_A}$, the partial transpose of the density matrix of the ground state, $\rho$\cite{horodecki2009quantum}. $\mathcal{N}_E=0$ if $\rho$ is separable, while $\mathcal{N}_E>0$ for an entangled state. For the icosahedron shell, $\mathcal{N}_E$ is computed by partitioning 12 vertices into two hemispherical region (one of them is highlighted as the blue shaded region in the inset of Fig.\ref{fig: entanglement}). Fig.\ref{fig: entanglement} (a) illustrates the entanglement is instantaneously generated for non-zero $J_\pm$.

\section{Discussion and Conclusion}
\label{sec:discussion}

In summary, we discover the magnetic quasicrystalline systems host multipolar degrees of freedom, as a result of the interplay of spin-orbit coupling and CEF splitting of the icosahedral point group symmetry. 
Despite intensive previous work on magnetic properties in quasicrystals, the possibility of multipolar degrees of freedom and their importance have never been pointed out. 
In our work, for the first time, we have shown that the noncrystallographic point group symmetry can generally host such  unconventional degrees of freedom and have studied their characteristics.   
Depending on a perfect icosahedral symmetry $I_h$ or five-fold symmetry $C_{5v}$, we classify the multipole characteristics for different values of $J$. Strikingly, we have shown that pure octupolar degrees of freedom emerges for  $J=7/2$ under $I_h$ symmetry.  In this case, the Kramers doublet has zero expectation values for both magnetic dipoles and quadrupoles, however, the rank 3 tensors which represent octupole degrees of freedom, have non-vanishing expectation values. Based on the symmetry transformation, we further clarify each component of the pseudospin-1/2 in terms of magnetic octupoles as discussed in Eq.\eqref{octupoles}. 
 On symmetry grounds, we also derive the spin exchange Hamiltonian with four independent parameters.
For antiferromagnetic Ising model, magnetic frustration leads to 72 degenerate states for a single icosahedron.
For a small but finite $J_\pm$, quantum fluctuations make a particular mixture of these degenerate states. It makes different but unique ground states for (anti-) ferromagnetic $J_\pm$, producing a finite entanglement even for arbitrary small  $J_\pm$. 
Depending on the inter-shell distances, possible macroscopic degeneracy and entanglement of octupoles would be an interesting future work. 
Also, the studies in the presence of $J_{\pm\pm}$ and $J_{\pm z}$, which do not preserve the total $\Sigma^{z}$, can be explored which we leave as a future work. 

Such octupolar degrees of freedom can be found in rare-earth based magnetic quasicrystals and approximants such as  Au-Al alloys, Cd-Mg alloys including rare earth ions and etc\cite{walter1987crystal,labib2020magnetic,kashimoto1998magnetic,watanabe2021topological,suzuki2021magnetism,deguchi2012quantum,watanuki2011pressure,sato2001magnetic, imura2023variation,yamada2019formation,PhysRevB.61.476,PhysRevB.100.054417,doi:10.1021/jacs.1c09954,hiroto2014sign}. 
However, many of the currently present magnetic quasicrystals have issues of intermediate valence and  some mixed sites between non rare-earth atoms. \cite{watanabe2021topological,watanuki2012intermediate,watanuki2011pressure,lehr2013enhanced, matsukawa2014valence, otsuki2016distributed, watanabe2016origin, hayashi2017observation, watanabe2021magnetism,hiroto2014sign,imura2023variation}. It makes imperfect symmetries,  allowing small deviation from perfect non-crystallographic point group symmetries. Nonetheless, one expects that advancements in chemical synthesis techniques could make the successful synthesis of finely controlled icosahedral quasicrystals\cite{watanuki2011pressure,lehr2013enhanced,yamada2019formation, nagaoka2023quasicrystalline}, and it may give us a chance to discover pure magnetic octupoles or even higher order multipolar degrees of freedom and their interesting physics.


Our work for the first time shows that the multipolar degrees of freedom naturally emerge in icosahedral quasicrystals. It breaks new ground in the magnetism of quasicrystals, and opens several interesting questions. One could explore magnetic quasicrystals searching for hidden phases, magnetic frustration induced long range entanglement such as spin liquids and non-Fermi liquids due to exotic Kondo effect\cite{kimura1986temperature,widom1988short, goldman2014magnetism, andrade2015non, matsukawa2016pressure, shi2018frustration,PhysRevLett.115.036403, tamura2021experimental}. 
Our study motivates to experimentally find new rare-earth icosahedral quasicrystals beyond conventional magnetism in periodic crystals. The field of magnetic quasicrystals is an interesting research area, and continued advancements in both experimental and theoretical studies lead us to discover new magnetic phenomena in quasicrystals.

\section{Methods}\label{sec:method}

To construct the CEF Hamiltonian, we used the simplified point charge model, where the point charge ligands are placed on each 12 verticies of the icosahedron surrounding the rare earth atom. For the charge impurity model where the icosahedral point group symmetry breaks down to the $C_{5v}$, we place the charge impurity on the ligand site placed on the local $z$-axis. To compute the Stevens parameter, we use the numerical values of the Stevens factors and radial integrals of rare-earth ions in \cite{EDVARDSSON1998230}. By block diagonalizing the CEF Hamiltonians, we generally search the possible doublet eigenspaces, and their possible multipolar degrees of freedom. We examine the vanishing of the lower magnetic or electric multipole operators to conclude the presence of the higher order multipolar degrees of freedoms. To construct the symmetry allowed spin Hamiltonian on the icosahedral quasicrystal, we apply the icosahedral point group symmetry, and time-reversal symmetry to the octupolar pseudospin operators---that is described in the Supplementary Materials in detail.

The icosahedral quasicrystal is constructed by the standard cut-and-project scheme described in the Supplementary Materials in detail. Considering the open boundary condition, we use the exact diagonalization method and symmetry ground to find the ground states of the Ising model. With quantum fluctuation, we analytically investigate the purely entangled ground state, and numerically calculate its entanglement negativity emergent on the single icosahedron shell by using the exact diagonalization method.


\textbf{Acknowledgments}

We thank Takanori Sugimoto and Taku J Sato for useful discussions. J.M.J and S.B.L. are supported by National Research Foundation Grant (No. 2021R1A2C1093060)).

\textbf{Competing interests}
The authors declare no conflict of interest.

\textbf{Author contributions}
J.M.J and S.B.L develop main idea of the project. J.M.J also produces the data and analyzes them. All authors contribute for writing the manuscript.

\textbf{Data availability}
The data that support the findings of this study are available from the corresponding author upon request.

\bibliography{my1}

\clearpage
\pagebreak

\renewcommand{\thesection}{\arabic{section}}
\setcounter{section}{0}
\renewcommand{\thefigure}{S\arabic{figure}}
\setcounter{figure}{0}
\renewcommand{\theequation}{S\arabic{equation}}
\setcounter{equation}{0}

\begin{widetext}

\begin{bibunit}
	\section*{Supplementary Material}
\renewcommand{\thefigure}{S\arabic{figure}}
\setcounter{figure}{0}
\renewcommand{\theequation}{S\arabic{equation}}
\setcounter{equation}{0}

\section{Crystal electric field Hamiltonian and Stevenes operator}
The crystal field Hamiltonian under the icosahedral point group symmetry would be written as $H_{CEF}=B_6(O_6^0-42O_6^5)$.
Here, $B_6=A_6\gamma_J\braket{r^6}$ is the Stevens coefficient obtained by the radial integral. $\gamma_J$ is the Stevens factor, and $r$ is the radial position. Especially, $A_6=-\frac{33}{100} \frac{q_0|e|}{R_0^7}$ where $q_0$ is the charge of the ligands and $R_0$ is the distance between ligands and central rare earth atom\cite{walter1987crystal,lindgard1975tables}. We focus on the angular parts which are $O_6^0$ and $O_6^5$, Stevens operators. They are given by\cite{walter1987crystal} 
\begin{align}
\label{stevenes}
&O_6^0=231J_z^6-105(3J(J+1)-7)J_z^4+(105J^2(J+1)^2-525J(J+1)+294)J_z^2 -5J^3(J+1)^3+40J^2(J+1)^2-60J(J+1) \\ &O_6^5=\frac{J_z(J_+^5+J_-^5)+(J_+^5+J_-^5)J_z}{4}
\end{align}

If the central rare earth atom is $\mbox{Yb}^{3+}$, the $4f$ electrons are the valance electron. It is known that $q_0$ would be positive for some materials such as Au-Al-Yb compounds. In such case, we have $A_6<0$. Furthermore, $\mbox{Yb}^{3+}$ has the total angular momentum $J=7/2$ with $\gamma_J=1.48\times 10^{-4}$, so we have $B_6<0$\cite{walter1987crystal,lindgard1975tables}. In this case the ground state sector of $H_{CEF}$, is Kramer doublet.

Once the icosahedral point group symmetry is broken down to the $C_{5v}$ symmetry due to the single point charge impurity placed on the $z$-axis, the additional diagonal Stevens operators, $O_2^0$ and $O_4^0$ could be appeared. They are defined by
\begin{align}
\label{stevenes}
&O_2^0=3J_z^2-J(J+1) \\ &O_4^0=35J_z^4-(30J(J+1)-25)Jz^2+3J^2(J+1)^2-6J(J+1)
\end{align}

\section{Local axes and symmetry allowed Hamiltonian}

We consider an icosahedron whose one of the 5-fold rotationally symmetric axes is the global $z$-axis. We define the local coordinate axes for the icosahedron as follows.
\begin{align}
\label{local axes}
& \vec{z}_1=(0,0,-1)\\
& \vec{z}_2=-\frac{1}{\sqrt{5}}(2,0,1)\nonumber \\
& \vec{z}_3=-\left(\frac{5-\sqrt{5}}{10},\sqrt{\frac{5+\sqrt{5}}{10}},\frac{1}{\sqrt{5}}\right)\nonumber \\
& \vec{z}_4=-\left(\frac{-5-\sqrt{5}}{10},\sqrt{\frac{5-\sqrt{5}}{10}},\frac{1}{\sqrt{5}}\right)\nonumber \\
& \vec{z}_5=-\left(\frac{-5-\sqrt{5}}{10},-\sqrt{\frac{5-\sqrt{5}}{10}},\frac{1}{\sqrt{5}}\right)\nonumber \\
& \vec{z}_6=-\left(\frac{5-\sqrt{5}}{10},-\sqrt{\frac{5+\sqrt{5}}{10}},\frac{1}{\sqrt{5}}\right)\nonumber
\end{align}
Note $\vec{x}_i=(0,1,0)\times\vec{z}_i$ and $\vec{y}_i=\vec{z}_i\times\vec{x}_i$. The other six coordinates are obtained by inversion symmetry. Specifically, the local $z$ and $x$ axes of $i\leftrightarrow 13-i$ sites are related by the inversion transformation. Note that $\vec{z}_i$ is the position vector of the site from the center of the icosahedron shell.

The symmetry allowed spin Hamiltonian with respect to the local coordinate axes is obtained by applying two mirror reflection symmetries, $M_1$ and $M_2$ whose normal vectors of the mirror planes are $\vec{n}_1=(0,1,0)$ and $\vec{n}_2=(1,0,\varphi)/\sqrt{1+\varphi^2}$. Since $\Sigma^\mu$, where $\mu=x,y,z$ are the pseudospin operators which are the rank 3 tensor operators, the unitary representation of the reflection transformation is given by $\pi$-rotation around these normal vectors of the mirror planes for angular momentum $j=3$, octupoles. Considering $U(M_{1(2)})$ as the unitary representation of the mirror reflection, $M_{1(2)}$, then the $T_m^{(3)}$ operators transform under the mirror reflections $M_{1(2)}$ as in Eq.\eqref{symmetrytransform}.
\begin{align}
\label{symmetrytransform}
&T_m^{(3)}\to U(M_{1(2)})T_m^{(3)} U^\dagger(M_{1(2)})=\sum_{m'}D_{1(2),m'm}^{(3)}T_{m'}^{(3)}
\end{align}
where $D_{1(2),m'm}$ are the Wigner D-matrix elements for the angular momentum $j$. Each of the spherical tensor operator is represented as the pseudospin operators in the Kramers doublet, as discussed in the main text. To be more specific, the Wigner D-matrices are given by
\begin{align}
\label{wignerD}
&D_{1,m'm}^{(3)}=\begin{pmatrix}  0 &0&0&0&0&0&1 \\0 &0&0&0&0&-1&0 \\0 &0&0&0&1&0&0 \\0 &0&0&-1&0&0&0 \\ 0 &0&1&0&0&0&0 \\  0 &-1&0&0&0&0&0 \\ 1 &0&0&0&0&0&0 \end{pmatrix} \\
\label{wignerD2}
&D_{2,m'm}^{(3)}=\begin{pmatrix}  \frac{-5-2\sqrt{5}}{25} &\frac{1}{5}\sqrt{\frac{3}{5}(7+3\sqrt{5})}&-\frac{1}{5}\sqrt{\frac{3}{2}(3+\sqrt{5})}&\frac{2}{5}&-\frac{1}{5}\sqrt{\frac{3}{2}(3-\sqrt{5})}&\frac{1}{5}\sqrt{\frac{3}{5}(7-3\sqrt{5})}&\frac{-5+2\sqrt{5}}{25} \\ \frac{1}{5}\sqrt{\frac{3}{5}(7+3\sqrt{5})}&\frac{-15-\sqrt{5}}{50}&-\frac{1}{5}\sqrt{3-\sqrt{5}}&\frac{\sqrt{6}}{5}&-\frac{1}{5}\sqrt{3+\sqrt{5}}&\frac{15-\sqrt{5}}{50}&\frac{1}{25}\sqrt{\frac{3}{2}}(5-3\sqrt{5}) \\-\frac{1}{5}\sqrt{\frac{3}{2}(3+\sqrt{5})} &-\frac{1}{5}\sqrt{3-\sqrt{5}}&\frac{1}{\sqrt{5}}&0&-\frac{1}{\sqrt{5}}&\frac{\sqrt{3+\sqrt{5}}}{5}&-\frac{1}{5}\sqrt{\frac{3}{2}(3-\sqrt{5})} \\ \frac{2}{5}&\frac{\sqrt{6}}{5}&0&-\frac{1}{\sqrt{5}}&0&\frac{\sqrt{6}}{5}&-\frac{2}{5} \\ -\frac{1}{5}\sqrt{\frac{3}{2}(3-\sqrt{5})}&-\frac{1}{5}\sqrt{3+\sqrt{5}}&-\frac{1}{\sqrt{5}}&0&\frac{1}{\sqrt{5}}&\frac{\sqrt{3-\sqrt{5}}}{5}&-\frac{1}{5}\sqrt{\frac{3}{2}(3+\sqrt{5})} \\  \frac{1}{5}\sqrt{\frac{3}{5}(7-3\sqrt{5})}&\frac{15-\sqrt{5}}{50}&\frac{\sqrt{3+\sqrt{5}}}{5}&\frac{\sqrt{6}}{5}&\frac{\sqrt{3-\sqrt{5}}}{5}&\frac{1}{50}(-15-\sqrt{5})&-\frac{1}{5}\sqrt{\frac{3}{5}(7+3\sqrt{5})} \\ \frac{1}{25}(-5+2\sqrt{5})&\frac{1}{25}\sqrt{\frac{3}{2}}(5-3\sqrt{5})&-\frac{1}{5}\sqrt{\frac{3}{2}(3-\sqrt{5})}&-\frac{2}{5}&-\frac{1}{5}\sqrt{\frac{3}{2}(3+\sqrt{5})}&-\frac{1}{5}\sqrt{\frac{3}{5}(7+3\sqrt{5})}&\frac{1}{25}(-5-2\sqrt{5}) \end{pmatrix}
\end{align} 
Note that the Hermitian and time-reversal invariant Hamiltonian possesses 5 independent parameters. Under the mirror reflection symmetry, $M_1$ given by Eqs.\eqref{symmetrytransform} and \eqref{wignerD}, we have $\Sigma^{\pm}_{4(5)}\to -\Sigma^{\mp}_{5(4)}$ and $\Sigma^{z}_{4(5)}\to-\Sigma^{z}_{5(4)}$, where the subscripts 4 and 5 are the site indices (Refer to Fig.\ref{fig: phase}). Hence, there are only four independent parameters, $J_{\pm\pm},J_{z\pm},J_{\pm}$ and $J_{zz}$ for the Hamiltonian invariant under the mirror reflection $M_1$. On the other hand, we note that Eq.\eqref{wignerD2} does not give any further constraints on the parameters of the Hamiltonian. 

Under the 5-fold rotation, the pseudospin operators transform as $\Sigma_\pm^{(i)}\to e^{\mp4\pi/5}\Sigma_\pm^{(j)}$. Thus, the bond dependent phase factors are added in order to make the Hamiltonian invariant under the 5-fold rotations. Let $\alpha_{ij}$ and $\beta_{ij}$ be the matrix of the additional phase factor for the interaction term between the $i$-th site and the $j$-th site\cite{gaudet2019quantum,patri2020theory} (See the main text for the definition of $\alpha_{ij}$ and $\beta_{ij}$). Fig.\ref{fig: phase} shows the five types of nearest neighbor bond orientations, red, green, blue, cyan, and black. Referring to the indices in Fig.\ref{fig: phase}, the bond-orientation dependent phase factor are given as,
\begin{align}
\label{phases}
&\alpha_{ij}=1, &\mbox{ if } (i,j) \mbox{ or }(j,i)\in\{(1,2),(4,5),(3,7),(6,10),(8,9),(11,12)\}, \\
&\alpha_{ij}=e^{i2\pi/5}, &\mbox{ if } (i,j) \mbox{ or }(j,i)\in\{(2,3),(4,7),(1,5),(6.9),(10,11),(8,12)\}, \nonumber \\
&\alpha_{ij}=e^{i4\pi/5}, &\mbox{ if } (i,j) \mbox{ or }(j,i)\in\{(1,6),(3,4),(2,8),(5,11),(9,10),(7,12)\}\nonumber, \\
&\alpha_{ij}=e^{-i4\pi/5}, &\mbox{ if } (i,j) \mbox{ or }(j,i)\in\{(1,3),(5,6),(7,8),(4,11),(2,9),(10,12)\} \nonumber, \\
&\alpha_{ij}=e^{-i2\pi/5}, &\mbox{ if } (i,j) \mbox{ or }(j,i)\in\{(1,4),(3,8),(7,11),(9,12),(5,10),(2,6)\} \nonumber.
\end{align}
For a given $\alpha_{ij}$, $\beta_{ij}=(\alpha_{ij}^*)^2$.
\begin{figure}
  \includegraphics[width=0.5\textwidth]{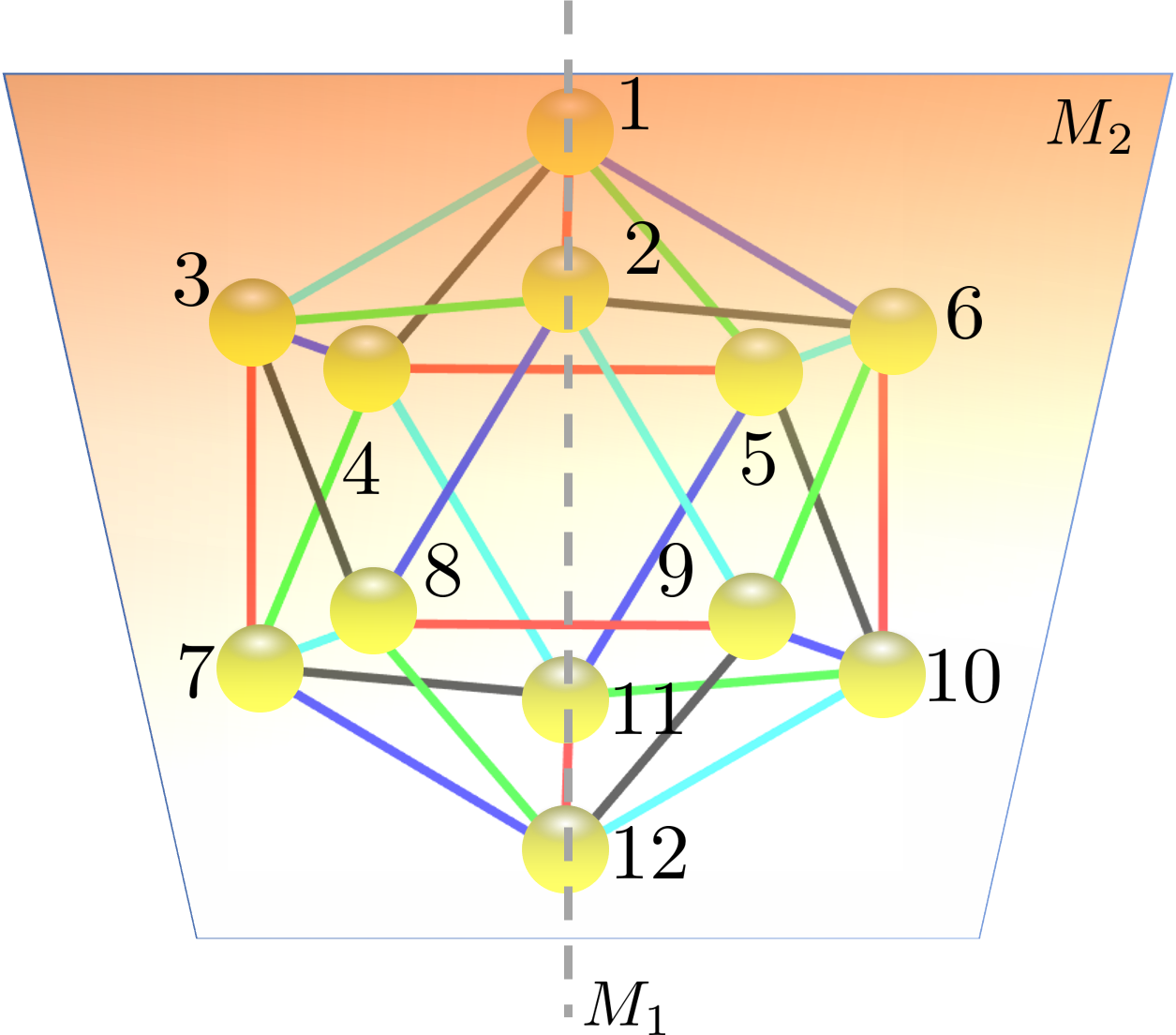}
  \caption{Five different bond orientations allow different phase factors. The red, green, blue, cyan and black colored nearest neighbor bonds have different phase factor matrix elements, $\alpha_{ij}$. The numbering of the icosahedron vertices indexes the sites shown as yellow spheres. Two mirror planes, $M_1$ and $M_2$, are shown as gray dashed line and orange plane, respectively. Note that $M_1$ swaps the sites 4 and 5, which are invariant under the $M_2$ reflection.}
  \label{fig: phase}
\end{figure}

\section{Icosahedron quasicrystal derived from cut-and-project scheme}
The cut-and-project scheme for constructing the icosahedral quasicrystal is introduced in this section\cite{kellendonk2015mathematics,jeon2021discovery}. The icosahedral point group symmetry is compatible in 6D space. Therefore, one can construct the icosahedral quasicrystal in 3D space as the lattice points descended from the 6D hypercubic lattice. In detail, let us consider the 6D hypercubic lattice $\mathcal{L}=\{ \boldsymbol{x} | \boldsymbol{x}=m_i \boldsymbol{e}_i, m_i\in\mathbb{Z}, 1\le i\le 6\}$, where $\boldsymbol{e}_i$ is a standard unit vector. The 6D space is decomposed by two projection maps $\boldsymbol{\pi}$ and $\boldsymbol{\pi}^{\perp}$. Each of them projects the lattice points in $\mathcal{L}$ onto the subspace of the quasicrystal (physical space) and its orthogonal complement subspace (perpendicular space), respectively. To produce the nontrivial quasicrystalline pattern, the physical space should have an irrational angle to the lattice surface. 
However, for such an irrational angle, the images of the projection of whole lattice points, $\boldsymbol{\pi}(\mathcal{L})$, densely cover the physical space. This violates the uniform discreteness of the definition of quasicrystals.
So one should choose a subset of $\mathcal{L}$, which is the (relatively) compact subset of the perpendicular space $\boldsymbol{\pi}^{\perp}$ and is often called the window, e.g. $K$.
Only if the image of $\boldsymbol{\pi}^{\perp}$ belongs to the window $K$, we project the lattice points. The resulting projection image in physical space is the discrete quasicrystalline lattice structure. As a standard choice of the window, $K=\boldsymbol{\pi}^{\perp}(\mathcal{W}(0))$, where $\mathcal{W}(0)$ is the Wigner-Seitz cell of the origin.
\begin{align}
\label{projection}
\boldsymbol{\pi} &=\begin{pmatrix} 0 & \tau & -1 & 0 & 1 & -\tau \\ \tau & -1 & 0 & -\tau & 0 & -1 \\ 1 & 0 & -\tau & 1 & -\tau & 0 \end{pmatrix} \\
\label{projectionperp}
\boldsymbol{\pi}^{\perp}&=\frac{1}{\sqrt{2+\tau}}\begin{pmatrix} 0 & -1 & -\tau & 0 & \tau & 1 \\ 1 & \tau & 0 & -1 & 0 & \tau \\ \tau & 0 & 1 & \tau & 1 & 0 \end{pmatrix}.
\end{align}
Then, the window $K$ becomes an icosahedron whose vertices are $\boldsymbol{\pi}^{\perp}$-projection images of the vertices of the Wigner-Seitz cell (See Fig.\ref{fig: window} (a).). The icosahedral quasicrystal (See Fig.\ref{fig: window} (b)) is given by the $\boldsymbol{\pi}$-projection of 6D hyper-cubic lattice points, whose image of $\boldsymbol{\pi}^{\perp}$ belongs to the window $K$ in Fig.\ref{fig: window} (a).
\begin{figure}
  \includegraphics[width=0.7\textwidth]{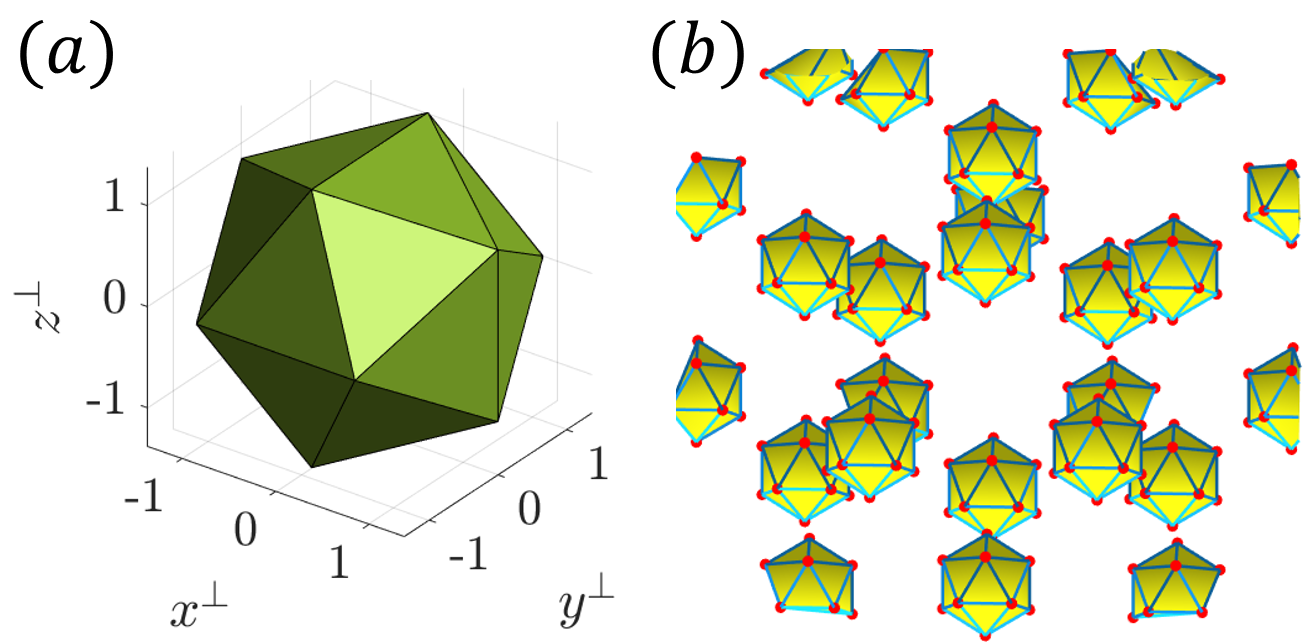}
  \caption{(a) Icosahedron window, $K$ of the cut and project scheme. (b) Icosahedral quasicrystal. The sites forming the icosahedron are shown only to avoid confusion. Red balls indicate the lattice sites. Nearest neighboring sites are connected by blue lines.}
  \label{fig: window}
\end{figure}

\section{Possible intershell magnetic orderings and frustrations}
This section shows an example of possible intershell magnetic states resulting from the intershell interaction terms. For each pair of nearest neighboring icosahedral shells, there are two pairs of nearest neighboring sites (see Fig. \ref{fig: intershellorder}). Fig.\ref{fig: intershellorder} shows an example of possible long-range intershell magnetic states regarding the octupolar degrees of freedom. In particular, Fig.\ref{fig: intershellorder} (a) shows the octupolar magnetic state when $J_{zz}$ is ferromagnetic for both the intra- and the intershell sites. Every icosahedron is ordered as one of the two ground states of the Ising model. On the other hand, Fig.\ref{fig: intershellorder} (b) shows the octupolar magnetic state when $J_{zz}$ is ferromagnetic for the intrashell sites but antiferromagnetic for the intershell sites. This gives rise to the \textit{inter-shell geometric frustration.} Note that the orange dashed lines representing inter-shell interactions form the triangles leading to intershell geometric frustration. Specifically, 12 icosahedron shells sit on the vertices of the inflated icosahedron shape in the icosahedral quasicrystal. Therefore, considering $\ket{FM_\pm}$ as two ferromagnetic Ising ground states of a single icosahedron shell, the antiferromagnetic interaction between the shells results in an antiferromagnetic Ising order on the inflated icosahedron in terms of $\ket{FM_\pm}$. This is how the geometrical frustration is created in the enlarged spatial scale in the icosahedral quasicrystal. 
Although we have given the two particular examples, depending on the intra-shell distances and their magnetic exchange couplings, it can stabilize different magnetic ground state case by case.
\begin{figure}
  \includegraphics[width=0.9\textwidth]{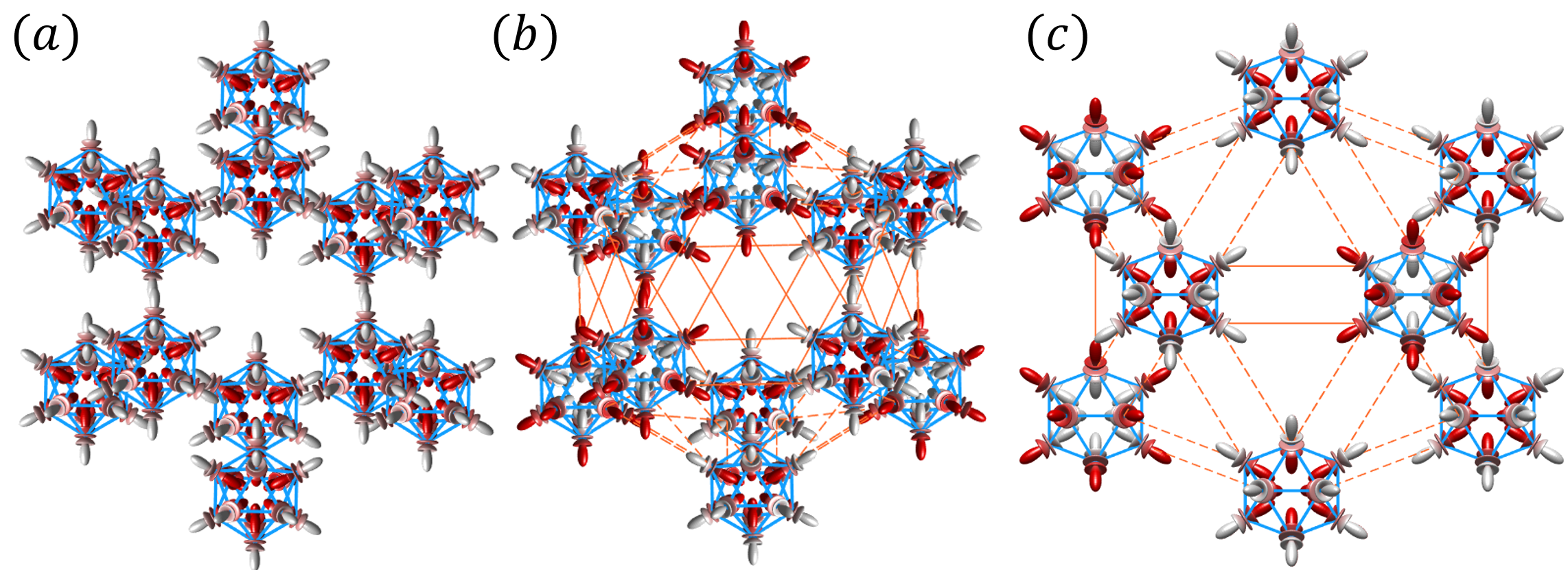}
  \caption{Examples of octupolar magnetic states in the presence of intershell interactions $J_{zz}$, where (a) both intra- and intershells are ferromagnetic, and (b) the intra-shell is ferromagnetic but the intershell is antiferromagnetic. The bonds between the nearest neighbors in the shell of an icosahedron are shown as blue lines. In (b), the nearest neighbor bonds between two adjacent icosahedron shells are shown as orange dashed lines. (c) Orthographic projection of (b) to emphasize that three icosahedron shells interact antiferromagnetically, leading to geometrical frustration.}
  \label{fig: intershellorder}
\end{figure}

\section{Derivation of the unique ground state for $|J_{\pm}|\ll J_{zz}$}
Here, we perturbatively derive the unique ground states emergent by small $J_\pm$ compared to $J_{zz}$. We apply the degenerate perturbation theory for 72-fold degenerate ground states for $J_{zz}>0$. Note that based on the symmetry ground, we can classify the 72 states into three groups, $\ket{\psi_{A_n}},\ket{\psi_{B_l}}$ and $\ket{\psi_{C_r}}$, where $1\le n,l\le 30$, while $1\le r\le 12$. Figs.\ref{fig: 72sup} (a,b,c) show the representative states in each group, $A,B,C$, respectively. Each group is generated by applying \textit{orientation-preserving} icosahedral symmetry operations to these states in Figs.\ref{fig: 72sup} (a,b) and (c), respectively. Every \textit{orientation-preserving} operations are in the maximal normal subgroup of the full icosahedral symmetry group, $I\trianglelefteq I_h$. Note $I$ is called as the icosahedral rotation group of order 60\cite{cornwell1997group}.
\begin{figure}
  \includegraphics[width=0.7\textwidth]{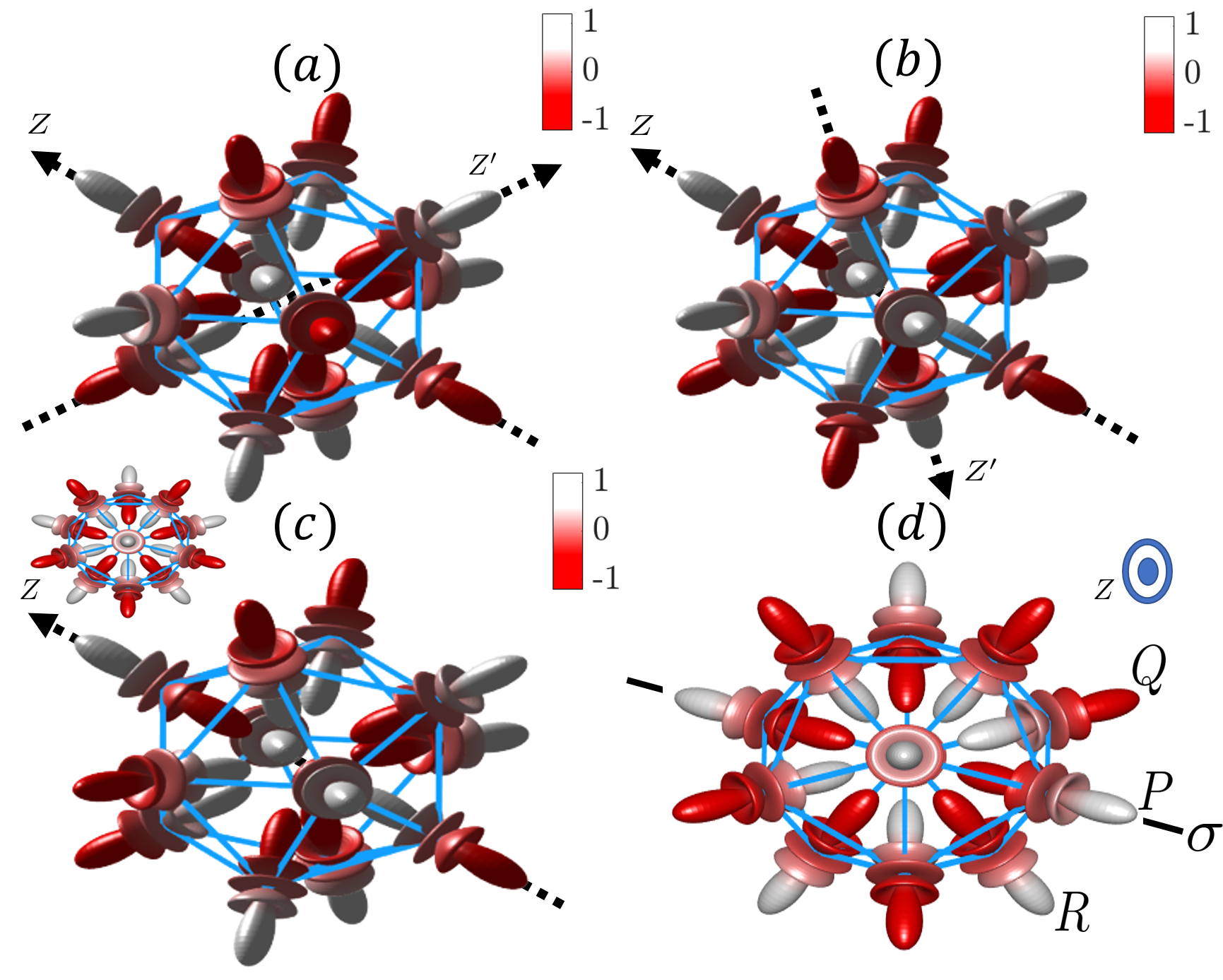}
  \caption{Classification of the 72 degenerate ground states on an icosahedron shell. (a,b) 60 degenerate states without 5-fold symmetry. The states admit the 2-fold symmetry operation, which transforms the $Z$ axis into the $Z'$ axis. (c) 12 degenerate states with 5-fold rotational symmetry around the $Z$ axis. The inset shows the view along the $Z$ axis to emphasise the 5-fold symmetry. (d) Top view along the $Z$ axis of the state shown in (a). There is a mirror plane, shown as $\sigma$, which includes the $Z$ axis and the site P, and which interchanges the octupoles at the sites Q and R, only under the mirror reflection of the site positions. As a result, the state shown in (a) is transformed into (b) and vice versa.}
  \label{fig: 72sup}
\end{figure}

First, the states shown in Figs.\ref{fig: 72sup} (a) and (b) represent the states without 5-fold rotational symmetry. Instead, they have another orientation-preserving icosahedral symmetry, which maps the $Z$ axis to the $Z'$ axis in Figs.\ref{fig: 72sup} (a) and (b) respectively. 
Since the icosahedral rotation group has the order, $|I|=60$, for each state in Figs.\ref{fig: 72sup} (a) and (b), we have 30 degenerate states generated by $I$, respectively. We denote them as $\ket{\psi_{A_n}}$ and $\ket{\psi_{B_l}}$, where $1\le n,l\le 30$. Note that the states in Fig.\ref{fig: 72sup} (a) and (b) are related by the mirror reflection of the sites with respect to the $\sigma$ plane depicted in Fig.\ref{fig: 72sup} (d). Here, $\sigma$ plane contains the $Z$ axis and the site P. This mirror reflection, $M_\sigma$ with respect to the $\sigma$-plane only exchanges the octupoles on the sites Q and R. Hence, the states, $\ket{\psi_{B_l}}$ are also obtained by applying the coset $IM_\sigma=\{gM_\sigma \vert g\in I  \}$ to the state in Fig.\ref{fig: 72sup} (a).

On the other hand, $\ket{\psi_{C_r}}$ are 12 degenerate states with 5-fold rotational symmetry (Refer to Fig.\ref{fig: 72sup} (c) and the inset representing the viewpoint along the rotational symmetry axis, $Z$ axis.). Since the state itself has 5-fold rotational symmetry, we have only 12 distinct orientation preserving transformations which give rise to the degenerate states in the group $C$. Hence, we may let the states in the group $C$ as $\ket{\psi_{C_r}}$, where $1\le r\le 12$.

Now, let us apply perturbative method to investigate the quantum fluctuation based on above groups. Take
\begin{align}
&H_\pm=J_\pm\sum_{\braket{i,j}}(\Sigma_i^+\Sigma_j^-+\Sigma_i^-\Sigma_j^+).
\end{align}
Then, we have 
\begin{align}
\label{zeroHpm}
&\braket{\psi_{A_n}|H_\pm|\psi_{A_m}}=\braket{\psi_{B_l}|H_\pm|\psi_{B_k}}=\braket{\psi_{C_r}|H_\pm|\psi_{C_s}}=0,
\end{align}
where $1\le n,m,l,k\le 30, 1\le r,s\le 12$. This is because $H_\pm$ has zero matrix element between two states related by the \textit{orientation preserving transformation}. Thus, one can write the block off-diagonal matrix representation of $H_\pm$ for the 72-fold degenerate states, say $[H_\pm]_{A,B,C}$ as
\begin{align}
\label{Hpm}
&[H_\pm]_{A,B,C}=\begin{pmatrix} 0 & T_{AB} & T_{AC} \\ T_{BA} & 0 & T_{BC} \\ T_{CA} & T_{CB} & 0 \end{pmatrix}
\end{align}
Here, the subscripts, $A,B$ and $C$ stand for the each groups. $T_{BA}=T_{AB}^{\dagger}$ is a $30\times30$ matrix, while $T_{AC}=T_{CA}^{\dagger}$ and $T_{BC}=T_{CB}^{\dagger}$ are $30\times12$ matrices. Here, each non-zero matrix element is $J_\pm$. We find that for each state in $A (B)$, there are 4, and 2 different states in $B (A)$ and $C$, respectively such that the matrix elements of $[H_\pm]_{A,B,C}$ is $J_\pm$. On the other hand, for each state in $C$, there are 5 different states in $A$ and $B$, respectively, and hence in total 10 states such that the matrix elements of $[H_\pm]_{A,B,C}$ is $J_\pm$.

One can use the graph, $G$ in Fig.\ref{fig: graph} to examine above facts. Here, the nodes of the graph $G$ represents to each state (red circle, green pentagram, and blue square represent the states in $A,B$ and $C$, respectively), and two nodes are connected by an edge if they admit nonzero matrix element of $[H_\pm]_{A,B,C}$. Thus, $[H_\pm]_{A,B,C}/J_\pm$ is the adjacency matrix of the graph $G$\cite{west2001introduction}. The graph $G$ is the decorated pentakis icosidodecahedron (See Fig.\ref{fig: graph}.). The pentakis icosidodecahedron (Fig.\ref{fig: graph} (a)) has 42 vertices in two different types depending on the local shape of the vertex, which are called pentagonal and hexagonal sites whose connectivity is 5 and 6, respectively\cite{bisztriczky2012polytopes}. Here, on the other hand, we term by the decorated pentakis icosidodecahedron, whose hexagonal sites are doubly occupied representing the states in $A$ (red circles) and $B$ (green pentagram), while the pentagonal sites of the pentakis icosidodecahedron are representing the states in $C$ (blue square) (See Fig.\ref{fig: graph} (b).). Hence, in the graph $G$, each node for the states in $A (B)$ are connected to the four nodes for the states in $B (A)$, and two nodes for the states in $C$, while each node for the states in $C$ are connected to the five nodes for the states in $A$ and $B$, respectively.
\begin{figure}
  \includegraphics[width=0.8\textwidth]{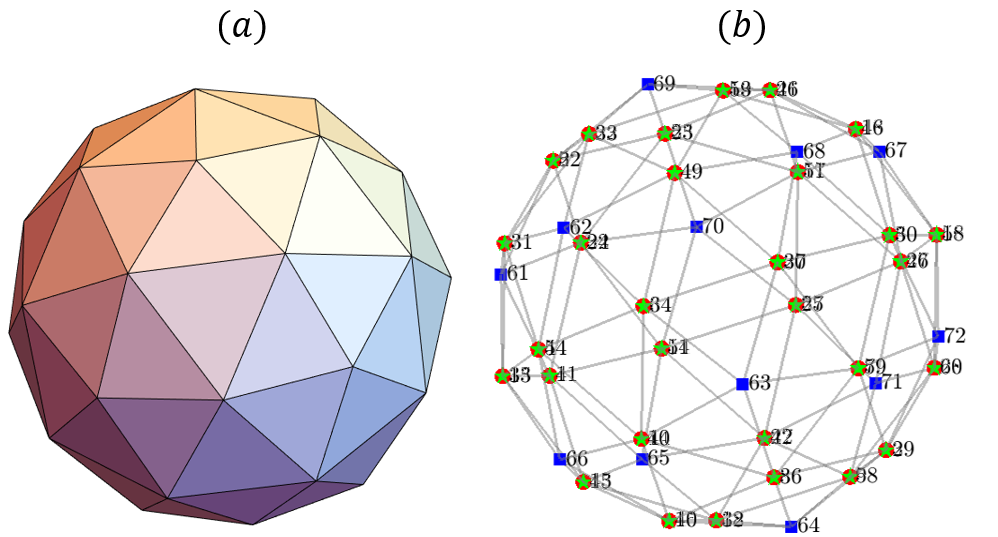}
  \caption{(a) Undecorated pentakis icosidodecahedron. There are 42 vertices, 120 edges, and 80 triangular faces. The 42 verticies are classified according to their two kinds of connectivity, 5 and 6. Each vertex having 5 (6) connectivity is called pentagonal (hexagonal) sites, respectively\cite{bisztriczky2012polytopes}. (b)  The graph whose adjacency matrix is $[H_\pm]_{A,B,C}/J_\pm$. We term this graph as decorated pentakis icosidodecahedron comprised of 72 nodes which represent the 72-fold degenerate antiferromagnetic ground states. The pentagonal vertices of the undecorated pentakis icosidodecahedron are doubly occupied by the states in $A$ and $B$ groups. If two states admit nonzero matrix element of $H_\pm$, then we have an edge depicted by the gray lines between two nodes representing each states. Red square, green pentagram, blue squares which have 6,6 and 10 connectivities, respectively represent the states in $A,B$ and $C$. The numberings of the nodes are the numberings of the state, where $1$ to $30$ are the states in $A$, $31$ to $60$ are the states in $B$, and $61$ to $72$ are the states in $C$.}
  \label{fig: graph}
\end{figure}

It allows us to write the general form of the ground state, $\ket{GS}$, as
\begin{align}
\label{GS}
&\ket{GS}=a\sum_{n=1}^{30}\ket{\psi_{A_n}}+b\sum_{l=1}^{30}\ket{\psi_{B_l}}+c\sum_{r=1}^{12}\ket{\psi_{C_r}},
\end{align}
where we have three real coefficients, $a,b$ and $c$ for $\ket{\psi_{A_n}},\ket{\psi_{B_l}}$ and $\ket{\psi_{C_r}}$, respectively. Note that $[H_\pm]_{A,B,C}$ is a real symmetry matrix, so we can assume without loss of generality that $a,b,c$ are real.
Then we have,
\begin{align}
\label{energy}
&E(a,b,c)=\braket{GS|H_\pm|GS}=240ab+120ac+120bc.
\end{align}
In addition, we have the normalization condition $N(a,b,c)=30a^2+30b^2+12c^2=1$.

The critical point is found by equating,
\begin{align}
\label{gradient}
&\nabla E(a,b,c)=\lambda\nabla N(a,b,c),
\end{align}
where $\lambda$ is the Lagrange multiplier. For $a=b=(1+\sqrt{6})c/5$, $E(a,b,c)$ is maximized, while for $a=-b$ and $c=0$, $E(a,b,c)$ is minimized. Each case corresponds to the unique ground state for the ferromagnetic and antiferromagnetic $J_\pm$. Remarkably, there is no degenerate ground state in either case. Thus, any small quantum fluctuation given by $H_\pm$ completely eliminates the degeneracy, by superposing 72 ground states of the antiferromagnetic Ising model.

\end{bibunit}
\end{widetext}

\end{document}